\documentclass[a4paper,12pt,twoside]{cpc-hepnp}

\usepackage{multicol}
\usepackage{graphicx}
\usepackage{booktabs}
\usepackage{amssymb,bm,mathrsfs,bbm,amscd}
\usepackage[tbtags]{amsmath}
\usepackage{lastpage}
\usepackage{xspace}
\usepackage{natbib}   
\usepackage{tabularx}
\mathchardef\Upsilon="7107
\def\Y#1S{\ensuremath{\Upsilon{(#1S)}}\xspace}

\newcommand{\nut}{\ensuremath{\nu_\tau}\xspace}

\newcommand{\BR}{\ensuremath{{\cal B}}\xspace}

\newcommand{\ee}{\ensuremath{e^+e^-}\xspace}

\newcommand{\Sew}{\ensuremath{S_{\rm EW}}\xspace}
\newcommand{\Gem}{\ensuremath{G_{\rm EM}}\xspace}
\newcommand{\tev}{\ensuremath{\mathrm{\,Te\kern -0.1em V}}\xspace}
\newcommand{\gev}{\ensuremath{\mathrm{\,Ge\kern -0.1em V}}\xspace}
\newcommand{\mev}{\ensuremath{\mathrm{\,Me\kern -0.1em V}}\xspace}
\newcommand{\kev}{\ensuremath{\mathrm{\,ke\kern -0.1em V}}\xspace}
\newcommand{\ev}{\ensuremath{\mathrm{\,e\kern -0.1em V}}\xspace}
\newcommand{\gevc}{\ensuremath{{\mathrm{\,Ge\kern -0.1em V\!/}c}}\xspace}
\newcommand{\mevc}{\ensuremath{{\mathrm{\,Me\kern -0.1em V\!/}c}}\xspace}
\newcommand{\gevcc}{\ensuremath{{\mathrm{\,Ge\kern -0.1em V\!/}c^2}}\xspace}
\newcommand{\mevcc}{\ensuremath{{\mathrm{\,Me\kern -0.1em V\!/}c^2}}\xspace}

\newcommand{\beq}{\begin{equation}}
\newcommand{\eeq}{\end{equation}}
\newcommand{\beqn}{\begin{eqnarray}}
\newcommand{\eeqn}{\end{eqnarray}}
\newcommand{\beqns}{\begin{eqnarray*}}
\newcommand{\eeqns}{\end{eqnarray*}}
\newcommand{\bitm}{\begin{itemize}}
\newcommand{\eitm}{\end{itemize}}

\newcommand{\amuhadLO}{\ensuremath{a_\mu^{\rm had,LO}}\xspace}

\newcommand{\ea}{{\em et al.}}
\newcommand\rs{\raisebox{1.5ex}[-1.5ex]}
\def\@citex[#1]#2{\if@filesw\immediate\write\@auxout{\string\citation{#2}}\fi
  \@tempcnta\z@\@tempcntb\m@ne\def\@citea{}\@cite{\@for\@citeb:=#2\do
    {\@ifundefined
       {b@\@citeb}{\@citeo\@tempcntb\m@ne\@citea
        \def\@citea{,\penalty\@m\ }{\bf ?}\@warning
       {Citation `\@citeb' on page \thepage \space undefined}}%
    {\setbox\z@\hbox{\global\@tempcntc0\csname b@\@citeb\endcsname\relax}%
     \ifnum\@tempcntc=\z@ \@citeo\@tempcntb\m@ne
       \@citea\def\@citea{,\penalty\@m}
       \hbox{\csname b@\@citeb\endcsname}%
     \else
      \advance\@tempcntb\@ne
      \ifnum\@tempcntb=\@tempcntc
      \else\advance\@tempcntb\m@ne\@citeo
      \@tempcnta\@tempcntc\@tempcntb\@tempcntc\fi\fi}}\@citeo}{#1}}

\def\@citeo{\ifnum\@tempcnta>\@tempcntb\else\@citea
  \def\@citea{,\penalty\@m}%
  \ifnum\@tempcnta=\@tempcntb\the\@tempcnta\else
   {\advance\@tempcnta\@ne\ifnum\@tempcnta=\@tempcntb \else
\def\@citea{--}\fi
    \advance\@tempcnta\m@ne\the\@tempcnta\@citea\the\@tempcntb}\fi\fi}
\newenvironment{myquote}
               {\list{}{\leftmargin0cm\indent}%
                \item\relax}
               {\endlist}
\newcommand\allFontSize{\footnotesize}
\newcommand\detailsSize{\allFontSize}
{\begin{myquote}\detailsSize}{\end{myquote}}

\begin{document}

\fancyhead[co]{\footnotesize G. L\'opez Castro: Recent progress on isospin breaking corrections}

\footnotetext[0]{Received January 2010}

\title{Recent progress on isospin breaking corrections and their impact on the muon $g-2$ value\thanks{Supported by Conacyt, M\'exico }}

\author{%
      Gabriel L\'opez Castro
}
\maketitle

\address{%
Departamento de F\'\i sica, Cinvestav, Apartado Postal 14-740, 07000 M\'exico, D.F M\'exico}

\begin{abstract}
We describe some recent results on isospin breaking corrections which are of relevance for predictions of the leading order hadronic contribution to the muon anomalous magnetic moment $\amuhadLO$ when using $\tau$ lepton data. When these corrections are applied to the new combined data on the $\pi^\pm\pi^0$ spectral function, the prediction for $\amuhadLO$ based on $\tau$ lepton data gets closer to the one obtained using $e^+e^-$ data.
\end{abstract}



\begin{multicols}{2}

\section{Introduction}

Currently, the accuracy of the Standard Model prediction of the muon anomalous magnetic moment $a_{\mu}=(g-2)/2$ is limited by the uncertainties of hadronic contributions \cite{amuhad, Davier-Beijing}. The dominant term in the leading order hadronic contribution $\amuhadLO$ and an important part of its associated uncertainty is provided by the $\pi\pi$ spectral function, which can be measured in $e^+e^-$ annihilations and in $\tau$ lepton decays (more details about their current status are given in the accompanying contribution by Michel Davier \cite{Davier-Beijing}). Owing to the isotopic properties of the electromagnetic and $\Delta S=0$ weak vector currents, the so-called Conserved Vector Current (CVC) hypothesis, the spectral functions themselves and their contributions to $\amuhadLO$ must be the same after isospin breaking (IB) corrections are appropriately applied to input data~\cite{Alemany98}.

 In recent years, a comparison of $e^+e^-$ and $\tau$ based  measurements of the $\pi\pi$ spectral functions in the timelike region, have shown large discrepancies for center of mass energies above the $\rho(770)$ resonance peak, beyond the size expected for IB corrections \cite{dehz02,dehz03}. The predictions for $\amuhadLO$ based on these two sets of data have been in  disagreement by more than $2\sigma$ \cite{dehz03,davier-pisa}. Moreover, the branching fraction for $\tau \to \pi\pi\nu$ predicted from $e^+e^-$ data corrected by IB effects was underestimated by more than $4\sigma$ with respect to the average of direct measurements \cite{dehz02, davier-pisa}. Given this `$e^+e^-$ vs $\tau$ discrepancy'\footnote{Another reason is that $e^+e^-$ data is more directly related to $\amuhadLO$ through the dispersion integral than $\tau$ decay data.}, it is believed that $\tau$ decay data does not provide at present a reliable determination of $\amuhadLO$ (currently, other useful contributions from $\tau$ decay data involve only the $2\pi$ and $4\pi$ channels \cite{Davier-Beijing}). Even though unidentified errors may be affecting $e^+e^-$ and/or $\tau$ decay data, understanding IB corrections becomes crucial to gain confidence about this important $2\pi$ contribution and, when consistency is achieved, to have a more precise prediction of $\amuhadLO$ from combined results.

In this contribution we summarize some recent results on the isospin breaking corrections that are relevant for understanding such discrepancies. As it is discussed in \cite{Davier-Beijing,nos09}, their application to the evaluation of $\amuhadLO$ from $\tau$ data leads to a value \cite{nos09} that is closer to $e^+e^-$-based calculations. The prediction of the $\tau \to \pi\pi\nu$ branching fraction based on the IB corrected $e^+e^-$ data also shows a reduced discrepancy with respect to results of direct measurements \cite{Davier-Beijing,nos09}. A preliminary version of this work appeared in ref. \cite{ib-novo}. In this new version we include a discussion of the complete set of IB corrections and we address some points that were unclear in some of our previous reports.

\section{IB corrections to the $\amuhadLO$ dispersion integral from $\tau$ data}

  The leading order hadronic contribution $\amuhadLO$ can be evaluated by using a combination of experimental data and perturbative QCD for the hadronic vacuum polarization (HVP) function of the photon.
 At low energies, where QCD does not provide a reliable calculation of Green functions, the HVP can be constructed as a sum over exclusive hadronic channels measured in $e^+e^-$ annihilation. The dispersion integral relating each exclusive $e^+e^- \to X^0$ channel to $\amuhadLO$ is:
\beq
\label{dis-int}
\amuhadLO~[X^0, e^+e^-]=\frac{\alpha^2}{3\pi^2}\vint_{4m_{\pi}^2}^{\infty}ds
\frac{K(s)}{s} R^{(0)}_{X^0}(s) \ ,
\eeq
where $R^{(0)}_{X^0}(s)$ is the ratio of hadronic $X^0$ to pointlike $\mu^+\mu^-$ {\it bare} cross sections \cite{amuhad, Davier-Beijing} in $e^+e^-$ annihilation at a center of mass energy $\sqrt{s}$. The behavior of the QED kernel $K(s)\sim 1/s$  \cite{br68}, enhances the low-energy contributions to $\amuhadLO$ in such a way that 91$\%$ of it comes from the energy region below 1.8 GeV and 73$\%$ is due to the $\pi\pi$ channel. Further details can be found in \cite{Davier-Beijing}.

  If isospin were an exact symmetry, we would be able to use in (\ref{dis-int}) the spectral functions measured in $\tau^- \to X^-\nu$ decays, where $X^-$ is the $(I=1, I_3=-1)$ isotopic partner of the $X^0$ state. We can define an isotopic analogue of the ratio $R^{(0)}_{X^0}(s)$ as follows (this quantity is related to the usual spectral function \cite{Davier-Beijing, nos09} by $R^{(0)}_{X^-}(s)=3v_{X^-}(s)$):
\beqn
\label{eq:sf}
   \frac{R^{(0)}_{X^-}(s)}{3}
   &=&
           \frac{m_\tau^2}{6\,|V_{ud}|^2}\,
              \frac{\BR_{X^-}}
                   {\BR_{e}}\,
              \frac{1}{N_X}\frac{d N_{X}}{ds}  \\
   & & 
              \times\,
              \left(1-\frac{s}{m_\tau^2}\right)^{\!\!-2}\!
                     \left(1+\frac{2s}{m_\tau^2}\right)^{\!\!-1}
              \, . \nonumber
\eeqn
In Eq.~(\ref{eq:sf}), $(1/N_X)dN_X/ds$ is the normalized invariant mass 
spectrum of the hadronic final state, and $\BR_{X^-}$ denotes 
the branching fraction of $\tau\to X^-(\gamma)\nut$ (throughout this paper, 
final state photon radiation is 
implied for $\tau$ branching fractions). For numerical purposes \cite{nos09}, we use for the $\tau$ lepton mass the value
$m_\tau=(1776.84\pm 0.17)\,{\rm MeV}$~\cite{pdg08}, and for the CKM matrix 
element $|V_{ud}|=0.97418\pm0.00019$~\cite{ckmfitter}, which assumes CKM unitarity. 
For the electron branching fraction we use 
$\BR_{e}=(17.818 \pm 0.032)\%$, 
obtained~\cite{rmp} supposing lepton universality. 

  In the presence of IB effects, the spectral function (\ref{eq:sf}) in $\tau$ decays must
be corrected by the factor $R_{IB}(s)/\Sew$, in order to be used into the dispersion integral (\ref{dis-int}). Therefore, it becomes convenient to introduce the shift in the dispersion integral:
\beqn
\label{shift}
\Delta \amuhadLO~[X^-, \tau] &=& \frac{\alpha^2}{3\pi^2}\vint_{4m_{\pi}^2}^{\infty}ds
\frac{K(s)}{s} R^{(0)}_{X^-}(s) \nonumber \\
&& \ \ \  \times \left[\frac{R_{\rm IB}(s)}{\Sew}-1 \right]
\eeqn
produced by the IB corrections. The short-distance electroweak radiative effects encoded in $S_{EW}$, which includes the re-summation of terms of $O(\alpha^n\ln^n(m_Z))$ and of $O(\alpha \alpha^n_s\ln^n(m_Z))$, lead to the correction 
$\Sew=1.0235\pm0.0003$~\cite{dehz02,marciano, bl90,erler04}; the quoted uncertainty is attributed to   neglected corrections of $O(\alpha \alpha_s/\pi)$ \cite{erler04}. This term provides the largest of IB effects in $\amuhadLO~[X^-, \tau]$, as it can be seen in Table 1. The remaining IB effects included in $R_{IB}(s)$ are discussed below.

 Hereafter we focus on the $X^-=\pi^-\pi^0$ channel of $\tau$ lepton decays. Beyond its rather large contribution to $\amuhadLO$, the precision attained in the measurement of the muon anomalous magnetic moment  requires that the $\pi\pi$ contribution be evaluated below the 1$\%$ accuracy, making crucial the reliable computation of IB corrections \cite{Alemany98,dehz02}. The $s$-dependent IB correction introduced in (\ref{shift}) is defined as:
\beq
\label{rib}
R_{\rm IB}(s)=\frac{{\rm FSR}(s)}{\Gem(s)}
              \frac{\beta^3_0(s)}{\beta^3_-(s)}
              \left|\frac{F_0(s)}{F_-(s)}\right|^2\,.
\eeq
The subscripts $i=0,\, -$ refer to the electric charge of the $2\pi$ system produced in $e^+e^-$ annihilation and in $\tau^-$ lepton decays, respectively. Each of the factors in $R_{\rm IB}(s)$ becomes unity in the limit of isospin symmetry, thus also $R_{\rm IB}(s)=1$ in this limit.

 In the Standard Model of quarks and leptons interactions, isospin symmetry is broken by the mass difference of $u$ and $d$ quarks, and by the effects of electromagnetic interactions. At the hadron level, the IB effects introduce some model dependence: hadronic matrix elements that are related by isospin symmetry, get modified in the presence of IB effects by photonic interactions and by the mass and width splitting of hadrons involved. Therefore, the usual procedure to test isospin symmetry predictions consist in comparing `bare' hadronic matrix elements obtained from experimental data by removing the effects of IB corrections. 

In the following we consider each of the energy-dependent factors that enter in $R_{\rm IB}(s)$ and quantify their effects in $\Delta \amuhadLO~[\pi\pi, \tau]$. The corresponding corrections induced in the branching fraction of $\tau \to \pi\pi\nu$, which is an independent test of these IB corrections, can be found in \cite{Davier-Beijing, nos09}.

\subsection{FSR and phase space corrections}

  The final state photonic corrections  to $e^+e^- \to \pi^+\pi^-$, ${\rm FSR}(s)$, and the ratio of pion velocities $\beta_0(s)/\beta_-(s)$,  are the best known corrections to be considered in Eq. (\ref{rib}). The FSR correction is computed using scalar QED and its expression is known 
\end{multicols}
\begin{table}[htb]
  \caption[.]{\label{tab:amu}
    Contributions to $\amuhadLO~[\pi\pi, \tau]$ 
    from the IB corrections discussed in section 2 and ref. \cite{nos09}.
    The twofold corrections in the second column correspond to results obtained using the GS \cite{gounaris} and KS \cite{ks} parametrization of pion form factors, respectively. For comparison, the last column, denoted as DEHZ03, contains the results of Ref. \cite{dehz02}.} 
\setlength{\tabcolsep}{0.0pc}
\vspace{-3mm} \small
\begin{tabularx}{\columnwidth}{@{\extracolsep{\fill}}lccc} 
\hline\noalign{\smallskip}
      & \multicolumn{2}{c}{$\Delta \amuhadLO[\pi\pi, \tau]$ ($10^{-10}$)} & \\
 \rs{Source}  & GS model & KS model & \rs{DEHZ03} \\
\noalign{\smallskip}\hline\noalign{\smallskip}
\Sew                &  \multicolumn{2}{c}{$-12.21\pm0.15$} & $-12.1\pm 0.3$ \\
$\Gem$              &  \multicolumn{2}{c}{$ -1.92\pm0.90$}  & $-1.0$\\
FSR                 &  \multicolumn{2}{c}{$+4.67\pm0.47$}   &  $+4.5$\\
$\rho$--$\omega$ interference
                    & $+2.80\pm 0.19$ & $+2.80\pm 0.15$  & $+3.5\pm0.6$\\
$m_{\pi^\pm}-m_{\pi^0}$ effect on $\sigma$
                    &  \multicolumn{2}{c}{$ -7.88$}  &$-7.0$      \\
$m_{\pi^\pm}-m_{\pi^0}$ effect on $\Gamma_{\rho}$
                    & $+4.09$ & $ +4.02$  & $+4.2$ \\
$m_{\rho^\pm}-m_{\rho^0_{\rm bare}}$ 
                   & $0.20^{+0.27}_{-0.19}$ & $ 0.11^{+0.19}_{-0.11}$ & $0.0\pm2.0$  \\
$\pi\pi\gamma$, electrom. decays
                    & $ -5.91\pm0.59$ & $-6.39\pm 0.64$ & $-1.4\pm1.2$\\
\noalign{\smallskip}\hline\noalign{\smallskip}
               & $-16.07\pm 1.22$ & $-16.70\pm 1.23$ \\
\rs{Total}             & \multicolumn{2}{c}{$-16.07\pm 1.85$} & \rs{$-9.3\pm2.4$}\\
\noalign{\smallskip}\hline
\end{tabularx}
\end{table}
\begin{multicols}{2}
analytically \cite{fsr}.  From Figure 1 we observe that the effects of these two corrections are important close to threshold and they vanish rapidly for increasing values of $s$. The phase-space factor is very accurate as it depends only on the pion masses. Instead, we have attributed a $\pm 10\%$ uncertainty (see Table 1)  to the contribution of FSR in $\Delta \amuhadLO~ [\pi\pi, \tau]$ to account for possible deviations from scalar QED. As it has been pointed out in \cite{Davier-Beijing}, KLOE \cite{muller} and BABAR \cite{ppg} measurements of $\pi^+\pi^-\gamma(\gamma)$ in electron-positron collisions support the validity of this hypothesis within the uncertainties quoted above.

\subsection{Long-distance correction}

  The definition of the long-distance photonic correction $\Gem$ to the photon-inclusive hadronic spectrum in $\tau \to \pi\pi\nu$ decay can be found elsewhere \cite{cen, ld-nos}. The virtual + real soft-photon corrections (which gives  an infrared-convergent result and that we have named model-independent corrections in previous works \cite{ld-nos, ib-novo}) of Refs. \cite{cen, ld-nos} are very similar numerically, despite the different pion-form factors used in both cases (resonance chiral model \cite{pich} and vector meson dominance \cite{nos05} model, respectively). This is an expected behavior  since $\Gem$ is defined from the ratio of radiatively-corrected and tree-level $2\pi$ spectra \cite{ld-nos, ib-novo}, thus the form factor dependences largely cancel.

The main difference between the calculations of Refs. \cite{cen} and \cite{ld-nos} stems from the regular part (which is infrared finite, and we call model-dependent contributions in previous works) of real photon emission, and it can be traced back to the model-dependent contribution to $\tau \to \pi\pi\nu\gamma$ involving the $\rho\omega\pi$ vertex \cite{ld-nos}. In practice, most of the experiments remove from their $\pi^-\pi^0$ spectrum, the events associated to the decay chain $\tau^- \to \pi^- \omega(\to \pi^0\gamma)\nu$, leaving the interferences of this with other $\tau$ lepton radiative amplitudes in the their $\pi\pi$ invariant mass distributions \cite{nos09}. In order to remain consistent, we also removed from our $\Gem$ correction the square of the radiative amplitude 
\begin{center}
\includegraphics[width=8cm]{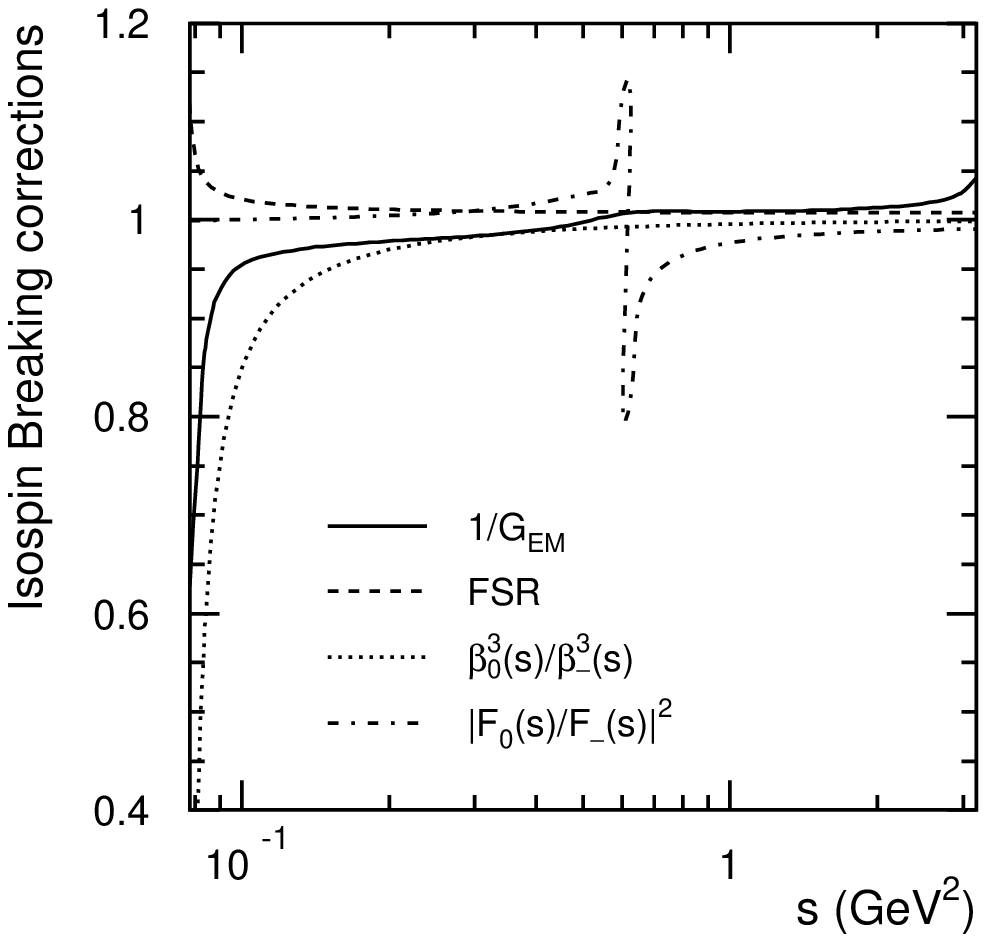}
\figcaption{\label{fig1} Energy-dependent IB corrections contained in $R_{\rm IB}(s)$, Eq. (\ref{rib}).}
\end{center}
involving the $\rho\omega\pi$ vertex \cite{nos09}. The resulting long-distance correction $\hat{G}_{\rm EM}$ gets closer to the one reported in ref. \cite{cen}. The inverse of  $\hat{G}_{\rm EM}$ is plotted as a solid line in Figure 1 and its effect in the dispersion integral (\ref{shift}) is  shown in the second row of Table 1. We have taken the difference between the effects of our $\hat{G}_{\rm EM}$ in $\Delta \amuhadLO~[\pi\pi, \tau]$ and the one of reference \cite{cen} (third column in Table 1) as an estimate of the uncertainty associated to model-dependence of the long-distance correction.

\subsection{IB effects in pion form factors}

  The last IB correction factor in Eq. (\ref{rib}), the ratio of the electromagnetic to the weak pion form factors, involves two sources of IB: $(a)$ a term that mixes the $I=1$ and $0$ components of the electromagnetic current which is driven by the $\rho-\omega$ mixing and, $(b)$ the mass and width difference of $\rho$ vector mesons which affect only the $I=1$ component of the form factors. We discuss this contribution in more detail since it represents the main change in $\Delta \amuhadLO[\pi\pi, \tau]$ with respect to previous evaluations of IB corrections.

  Under the above considerations, the pion form factors can be written as \cite{ib-novo, nos09}
\beqn
 F_{0}(s)\! &=&\! f_{\rho^0}(s)\left[1+\delta_{\rho \omega} 
   \frac{s}{m_{\omega}^2-s-im_{\omega}\Gamma_{\omega}(s)} 
                                    \right]\,,\\ 
 F_{-}(s)\! &=&\! f_{\rho^-}(s)\,,
       \eeqn
where the complex parameter $\delta_{\rho \omega}$ represents the strength of the $\rho-\omega$ mixing, and $f_{\rho^{0,-}}(s)$ denote\footnote{These form factors are normalized to unity when $s=0$.} the $I=1$ parts of the pion form factors which are dominated, below $\sqrt{s} \leq 1$ GeV, by the $\rho(770)$ vector  meson. 

There are different parametrizations of the form factors $f_{\rho^{0,-}}(s)$ in the literature which are inspired by different models \cite{maltman09} of the $\rho$ meson propagator. However, one would expect  that their ratio in Eq. (\ref{rib}) is relatively less sensitive to a particular model. Just for comparison, we adopt two commonly used phenomenological formulae: the Gounaris-Sakurai (GS) \cite{gounaris} and the K\"uhn-Santamaria (KS) \cite{ks} parametrizations.
Consequently, the corrections induced in $\Delta \amuhadLO~[\pi\pi, \tau]$ by the IB parameters in the pion form factors are quoted as two separate values in the second column of Table~1.

In the following we discuss the different sources of IB in formulae (5) and (6):

$\bullet$ {\bf Strength of $\rho-\omega$ mixing}: to obtain $\delta_{\rho\omega}$ we have fitted \cite{nos09} Eq. (5) with the GS and KS parametrizations to the $e^+e^-$ data in the full energy range available and we have included the effects of higher $I=1$ resonances in $F_{0}(s)$. This approach differs from other recent determinations of the $\rho-\omega$ mixing strength which obtain $\delta_{\rho\omega}$ from fits to \ee data below 1 GeV for a wider class of $F_0(s)$ models \cite{maltman09}. As a result of our fits reported in \cite{nos09}, we get for the strength and phases of the $\rho-\omega$ mixing parameter:
$|\delta^{\rm GS}_{\rho\omega}|=(2.00\pm 0.06)\times 10^{-3}$, 
       ${\rm arg}(\delta^{\rm GS}_{\rho\omega})=(11.6 \pm 1.8)^\circ$, and
       $|\delta^{\rm KS}_{\rho\omega}|=(1.87\pm 0.06)\times 10^{-3}$, 
       ${\rm arg}(\delta^{\rm KS}_{\rho\omega})=(13.2\pm 1.7)^\circ$.
In both cases, GS and KS, we use an energy-dependent absorptive part of the $\rho$ meson propagator given by $-i\sqrt{s}\Gamma_{\rho^{0,-}}(s)$. Contrary to claims raised in a recent paper \cite{maltman09}, we do not find a strong model-dependence of the $\delta_{\rho\omega}$ mixing parameter.
 
$\bullet$ {\bf Width difference of $\rho^{\pm}-\rho^0$ mesons} 
The energy-dependent decay widths of neutral and charged $\rho$ mesons cannot be measured in an independent way with the accuracy required  to estimate their effects in Eq. (3).  Thus, the width difference $\Delta \Gamma_{\rho} =\Gamma_{\rho^{\pm}} -\Gamma_{\rho^0}$ must be computed from the total widths which are defined as a sum over their exclusive decay channels \cite{wdiff}. A simple counting of decay channels of charged and neutral $\rho$ mesons give \cite{wdiff} 
\beqn
\Delta \Gamma_{\rho} &=&\Gamma[\rho^{\pm}\to \pi^{\pm}\pi^0(\gamma)] - \Gamma[\rho^0\to \pi^{+}\pi^-(\gamma)] \nonumber \\
&& \ \ \ -0.08\ {\rm MeV},
\eeqn
where the first two terms include the photon inclusive rates into two pions ($\pi\pi(\gamma_{\rm soft})+\pi\pi\gamma$). The last numerical term in Eq. (7) accounts for the rather small difference of the remaining decay widths \cite{pdg08} ($\pi\gamma, \eta\gamma, l^+l^-, \cdots$).  

The $2\pi$ photon inclusive decay rates, first line in (7), were calculated including the virtual plus real photon radiative corrections in \cite{wdiff}. We include  its energy-dependence in the following form: 
\beq
\label{deltag}
\Delta \Gamma_{\pi\pi(\gamma)}\! =\!\dfrac{g_{\rho\pi\pi}^2\sqrt{s}}{48\pi} \left[ \beta_-^3(s)(1+\delta_-) 
 -\beta_0^3(s)(1+\delta_0)\right] 
\eeq
where $g_{\rho\pi\pi}$ denotes the $\rho\pi\pi$ coupling and $\delta_{-,0}$ contains the effects of photonic radiative corrections with real photons of all energies. 
Eq. (\ref{deltag}) gives $\Delta \Gamma_{\pi\pi(\gamma)}= (-0.76\pm 0.08)$ MeV at $\sqrt{s}=m_{\rho}$, which can be compared with a previous estimate, $\Delta \Gamma[\rho \to \pi\pi(\gamma)]=(+0.49\pm 0.58)$ MeV \cite{Alemany98}, which was obtained by including only the effects of hard real photon emission \cite{singer}. The $\pm 10\%$ uncertainty added to our result for the width difference is estimated from the difference observed between our predicted branching fraction  for $\rho^0 \to \pi^+\pi^-\gamma$ \cite{wdiff} and its measured value~\cite{pdg08}.
\begin{center}
\includegraphics[width=7cm, angle=270]{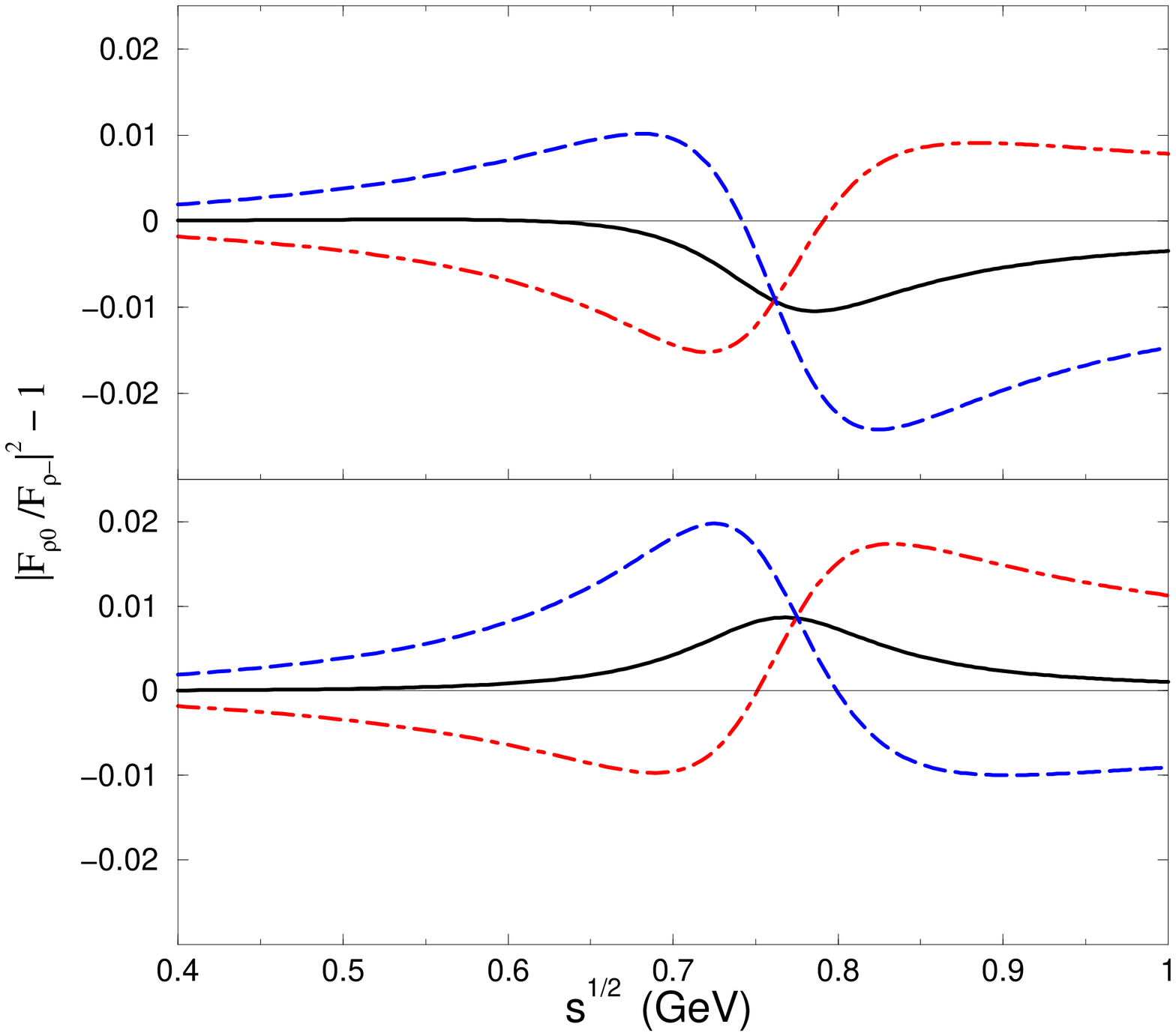}
\figcaption{\label{fig2} Comparison of the ratio of $I=1$ components of pion form factors: (upper plot) our results \cite{wdiff}, and (lower plot) the results of ref. \cite{dehz02}. The dashed, solid and dashed-dotted lines corresponds to $m_{\rho^{\pm}}-m_{\rho^0}=(+1, 0, -1)$ MeV. }
\end{center}
  As it can be seen from a comparison of the second and third columns in Table 1, the width difference (which we call `$\pi\pi\gamma$, electrom. decays') induces the biggest change in $\amuhadLO$ compared to results of  previous estimates \cite{Alemany98}. Just to emphasize the origin of this important change, in Figure 2 we compare the ratio of $I=1$ components of our form factors \cite{nos09, ib-novo} and the ones used in previous calculations \cite{dehz02}, for three different values of  the  mass difference: $m_{\rho^+}-m_{\rho^0}= (+1, 0, -1)$ MeV. The clear difference near the $\rho$ resonance peak, produces the large change in the IB effects due to $\pi\pi\gamma$ electromagnetic decays.

$\bullet$ {\bf Mass difference of $\rho^{\pm}-\rho^0$ mesons}. In previous analysis of the IB effects on $\Delta \amuhadLO$ \cite{dehz02} it was assumed that the charged and neutral $\rho$ mesons were degenerated, namely $\delta m_{\rho} \equiv m_{\rho^\pm}-m_{\rho^0}=(0\pm 1)$ MeV. An IB effect in the $\rho$ meson mass difference arises from the self-energy contribution generated by the $\rho^0-\gamma$ mixing term, which affects only the neutral $\rho$ meson mass by $m_{\rho^0}-m_{\rho^0_{\rm bare}} \approx 3\Gamma(\rho^0 \to e^+e^-)/(2\alpha)=1.45$ MeV~\cite{nos09}. When we remove this IB effect from the value $m_{\rho^\pm}-m_{\rho^0}=(-0.4\pm 0.9)$ MeV measured by KLOE \cite{kloe+0} from the Dalitz plot analysis of $\phi \to \pi^+\pi^-\pi^0$, we find $\delta m_{\rho} = (1.0\pm 0.9)$ MeV \cite{nos09}. We use this mass splitting in our evaluations shown in Table~1. 
\begin{center}
\includegraphics[width=7cm]{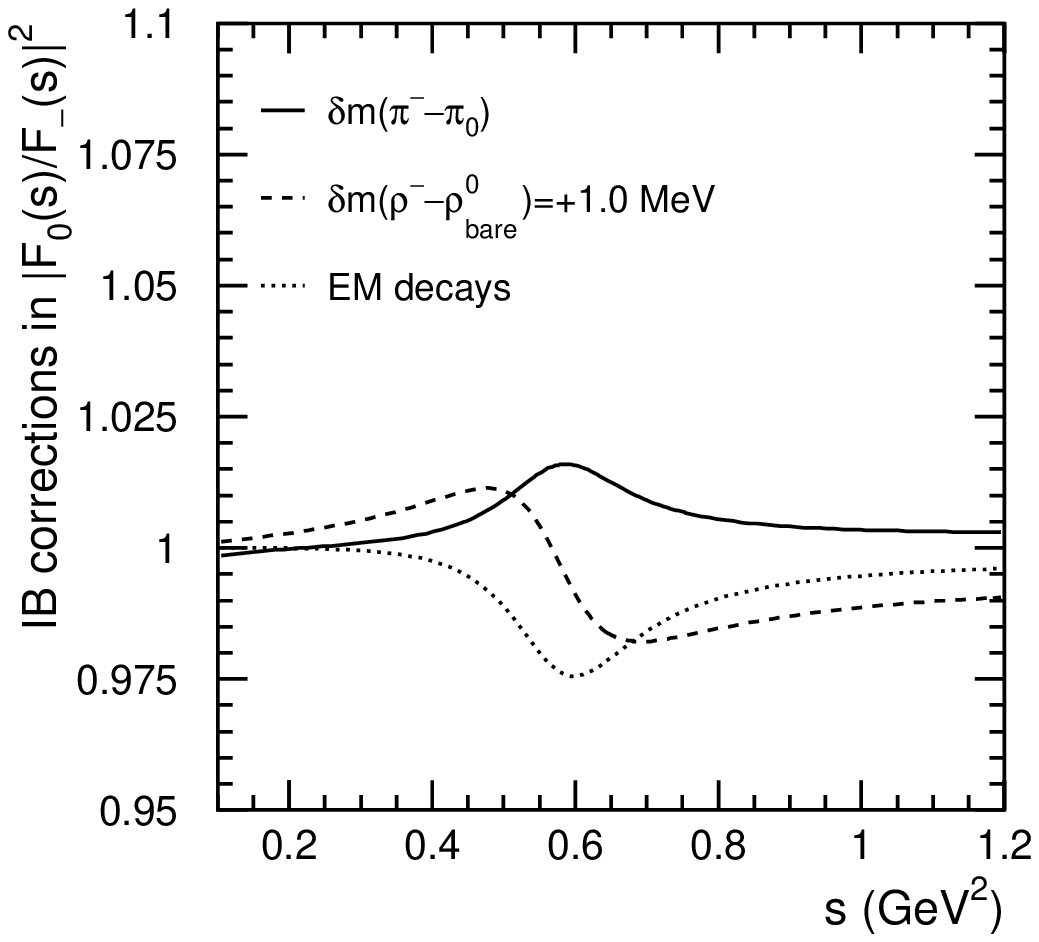}
\figcaption{\label{fig3} Contributions of mass and width differences to the ratio of $I=1$ components of form factors.}
\end{center}
 The IB effect produced in $\amuhadLO~[\pi\pi, \tau]$ by the $\rho$ mass difference is shown in Table 1 for the KS and GS parametrizations. As it can be observed, this effect gives a rather small contribution.
 
 In Figure 3 we plot separately the IB corrections in the ratio of the $I=1$ components of the pion form factors discussed in subsection 2.3, while their combined effects including the $\rho-\omega$ mixing term is represented with a dashed-dotted line in Figure 1. The IB effects that stems from the ratio of pion form factors are important close to the resonance peak, making $\amuhadLO~[\pi\pi, \tau]$ particularly sensitive to these corrections. The effect of the pion mass difference that affects the $\rho$ meson decay widths is shown in the 6th. row of Table 1 and with a solid line in Figure 3. Interestingly, the effects of the pion mass difference and of the $\pi\pi\gamma$ electromagnetic decays partly cancel each other as it can be observed from Table 1 and Figure~3.

\section{Conclusions}

  Isospin breaking (IB) corrections are of great relevance to improve the accuracy and gain confidence on the Standard Model  prediction of the leading order hadronic contribution $\amuhadLO$ to the muon anomalous magnetic moment. These corrections are also important in view of future and more precise measurements of  the muon anomalous magnetic moment \cite{roberts}. We have presented in this work, a summary of some recent results about these IB corrections.

 As it can be concluded from our results summarized in Table 1, the new IB corrections produce the change $\Delta \amuhadLO~[\pi\pi, \tau]=(-16.07\pm 1.85)\times 10^{-10}$ which is larger by $-6.8\times 10^{-10}$ units when compared to results used previously in \cite{dehz02}. The main change in the new corrections is due to the effect of the $\rho^{\pm}-\rho^0$ width difference \cite{wdiff}, which quantifies an important IB correction near the resonance of the $\pi\pi$ system. We have calculated, in two commonly used phenomenological models \cite{gounaris, ks}, the effects of IB corrections that are important around the $\rho$ resonance region, and have found that the model-dependence of pion form factors is not very important. 

The new IB corrections get closer the results of $\amuhadLO~[\pi\pi]$ based on \ee and $\tau$ lepton data (see \cite{Davier-Beijing, nos09}). These corrections also affect the prediction of the $\tau \to \pi\pi\nu$ branching fraction obtained from \ee data via the isospin symmetry.  As it was discussed in \cite{Davier-Beijing, nos09}, the large discrepancy  observed in previous comparisons of CVC predictions and direct measurements of this observable \cite{davier-pisa} is also reduced to an acceptable level after the new IB corrections are applied. It is very appealing that the new IB corrections reduce simultaneously the different manifestations of the so-called $e^+e^-$ vs. $\tau$ lepton discrepancy.

\

\acknowledgments{I would like to thank Changzheng and his colleagues for their invitation to this workshop and very kind hospitality. The results presented here are a summary of the work done  with M. Davier, A. H\"ocker, B. Malaescu, X. Mo, G. Toledo S\'anchez, P. Wang, C.Z. Yuan and Z. Zhang. I am very grateful to them for an enjoyable collaboration and very clarifying discussions.}

\end{multicols}

\vspace{-2mm}
\centerline{\rule{80mm}{0.1pt}}
\vspace{2mm}

\begin{multicols}{2}

\end{multicols}

\clearpage

\end{document}